Original Article

# Glycemic Variability Before And After Hypoglycemia Across Different Timeframes In Type 1 Diabetes With And Without Automated Insulin Delivery

*Glycemic variability related to hypoglycemia in T1D*

**Ahtsham Zafar[1], Abiodun Solanke[1], Dana M. Lewis[2], and Arsalan Shahid[1]**

1. CeADAR – Ireland's Centre for AI, University College Dublin, D04 V2N9 Dublin, Ireland
2. OpenAPS, Seattle, WA 98101, USA

**Ahtsham Zafar, MSc**
Data Scientist at CeADAR
Ireland's Centre for AI at University College Dublin
D04 V2N9 Dublin, Ireland
Email: ahtsham.zafar@ucd.ie
ORCID: 0000-0001-9691-2183

**Abiodun Solanke, PhD**
Data Scientist at CeADAR
Ireland's Centre for AI at University College Dublin
D04 V2N9 Dublin, Ireland
Email: abiodun.solanke@ucd.ie
ORCID: 0000-0002-6082-893X

**Dana M. Lewis, BA**
Independent Researcher
#OpenAPS
Seattle, WA, USA
Email: Dana@OpenAPS.org
ORCID: 0000-0001-9176-6308

**Arsalan Shahid, PhD, MBA (Corresponding Author)**
Technology Solutions Lead at CeADAR
Ireland's Centre for AI at University College Dublin
D04 V2N9 Dublin, Ireland
Email: arsalan.shahid@ucd.ie
ORCID: 0000-0002-3748-6361

## Abstract

**Background**

Managing Type 1 diabetes (T1D) aims to optimize glucose levels within the target range while minimizing hyperglycemia and hypoglycemia. Exercise presents additional challenges due to complex effects on glucose dynamics. Despite advancements in diabetes technology, significant gaps remain in understanding the relationship between exercise, glycemic variability (GV), and hypoglycemia in both automated insulin delivery (AID) and non-AID users. Additionally, limited research explores the temporal progression of GV before and after hypoglycemia and the impact of long-duration episodes on glucose recovery.

**Methods**

This study analyses the Type 1 Diabetes and Exercise Initiative (T1DEXI) dataset, assessing GV, hypoglycemia, gender, and exercise interactions in AID (n=222) and non-AID (n=276) users.. The study examined patterns of glycemic variability metrics like time below range (TBR) surrounding hypoglycemia events, focusing on the 48 hours before and after these events. We further assess the impact of different hypoglycemia levels (41–50 mg/dL, 51–60 mg/dL, and 61–70 mg/dL) on post-event glucose stability.

**Results**

Glycemic variability increased before and after hypoglycemia up to 48 hours in both AID and non-AID users, with statistically significant differences in GV metrics. TBR elevation persisted across all groups, peaking around hypoglycemic episodes. Notably, females using AID achieved significantly improved glucose stability compared to non-AID females, a larger within-group difference than that observed in males. Individual-level AID analyses revealed that long-duration hypoglycemia episodes (>40 minutes) resulted in prolonged TBR elevation, suggesting a slower recovery period despite AID intervention.

**Conclusion**

GV trends may aid in predicting hypoglycemia over extended time periods. Integrating GV patterns into AID systems could improve glucose stability and mitigate hypoglycemia cycles, especially with possible evaluation of hypoglycemia duration. Future research should explore hormonal influences (e.g., menstrual cycle effects) and inter-individual variability for optimised individual diabetes management.

**Introduction**

The management of type 1 diabetes has increasingly focused on maintaining glucose levels within a target range, commonly assessed through time in range (TIR), time above range (TAR), and time below range (TBR). Hypoglycemia, or low blood glucose levels, can arise from various situations, including mistimed insulin administration, excessive insulin dosing, increased physical activity, reduced meal absorption, or heightened insulin sensitivity[1]. Among these, physical activity and exercise present significant challenges in managing glucose levels due to their complex and unpredictable impact on glycemic outcomes[2]. Consequently, datasets capturing exercise-related glucose and insulin dosing data have become a focal point of research. One such dataset is the Type 1 Diabetes and Exercise Initiative (T1DEXI) dataset, which provides comprehensive continuous glucose monitoring (CGM) data across a diverse cohort of individuals with Type 1 diabetes, including a diversity of insulin delivery, demographic factors, and exercise routines[3].

The T1DEXI dataset has previously been characterized extensively in the literature[4], including the relationship between daily step count and glucose metrics[5] and the risk of hypoglycemia during and after exercise in individuals with impaired awareness of hypoglycemia[6]. Other notable studies include predicting hypoglycemia risk around exercise[7], an analysis of diet and glycemic outcomes[8], and applying machine learning techniques to classify unstructured exercise activities[9]. Further studies[10–14] have contributed to our understanding of glucose variability (GV) and hypoglycemia in the context of exercise, diet, and other factors from T1DEXI.

Despite significant progress in characterizing glycemic outcomes, critical gaps remain in understanding the interplay between exercise and glycemic variability (GV) among users of automated insulin delivery (AID) and non-AID systems (including pump and multiple daily injections (MDI) users). Additionally, the role of hypoglycemia in influencing GV has not been comprehensively examined across the population and individual levels in the T1DEXI dataset, although we have initially evaluated this in other large diabetes datasets of open-source AID users[15]. Addressing these gaps, this study leverages the T1DEXI dataset to perform both population-level and subgroup analyses, focusing on key metrics such as TIR, TBR, TAR, standard deviation (SD), mean glucose levels, coefficient of variation (CV), and the J-Index. We hypothesize that understanding the progression of glycemic variability related to periods of hypoglycemia will provide actionable insights for optimizing diabetes management technologies, including the efficacy of both AID and non-AID technologies. To this end, our study makes two primary contributions: (1) a comprehensive comparison of glycemic variability and hypoglycemia patterns between AID and non-AID users in the T1DEXI dataset, accounting for demographic factors and exercise types, and (2) a detailed analysis of glycemic variability progression related to hypoglycemia at both population and individual levels for commercial AID and non-AID systems.

## Methods

### Study Population

The T1DEXI dataset, derived from a real-world study on at-home exercise, includes adults with Type 1 diabetes (n=497) who were randomly assigned to complete structured exercise sessions over four weeks[4]. Previous research on T1DEXI has thoroughly characterised CGM data, structured exercise reports, demographic information, and raw insulin dosage data from insulin delivery devices[4,5].

### Data preparation and categorisation

For each participant, the T1DEXI dataset includes labels indicating the type of insulin pump or dosing technique used, including existing labels for AID or non-AID therapies. To ensure the accuracy of the existing labels, raw insulin dosing data were cross-checked for expected automated insulin adjustments (e.g., increased basal rate changes, temporary basal rates, or small boluses) to confirm that AID-labelled devices were used in automated rather than manual mode. Non-AID includes individuals using open-loop systems (including those with low glucose suspend features but without the ability to increase insulin for hyperglycemia), insulin pumps without automated insulin delivery functionality, or multiple daily injections (MDI). Figure A1 illustrates the total days of glucose data available across AID (n=222) and non-AID (n=276) categories. Based on these classifications, each subgroup was analysed independently.

### Glucose analysis metrics and statistical tests

Descriptive statistics were calculated, including mean, minimum, maximum, and quartiles, as well as standard deviation (SD). Additional glycemic outcome metrics calculated include TIR (70–180 mg/dL), TBR (<70 mg/dL), and TAR (>180 mg/dL). We employed the J-index, Low Blood Glucose Index (LBGI), High Blood Glucose Index (HBGI), and Coefficient of Variation (CV) to assess glycemic variability at different timeframes.

To perform a comparative analysis of glycemic variability between AID and non-AID users, we performed a variety of statistical tests including the Shapiro-Wilk (SW) test to test for normality in the data distributions. Additionally, the Z-test, Kolmogorov-Smirnov (KS) test and Mann-Whitney U (WMU) test were applied to compare the glucose data distributions among subgroups, including male vs. female and AID vs. non-AID users.

## Experimental workflow for population-level analysis of glycemic variability related to hypoglycemia

For population-level analysis, we adopted the experimental workflow from our previously published work[15] for assessing hypoglycemia through a lens of glycemic variability and applied it to AID and non-AID users in the T1DEXI dataset.

Hypoglycemia levels are categorised into three distinct ranges: 41–50 mg/dL, 51–60 mg/dL, and 61–70 mg/dL. (*Previous work[15] compared these ranges to "Level 1" (<70 mg/dL but >54 mg/dL) and "Level 2" (between 40 and <54 mg/dL) categories of hypoglycemia[16] and found no significant differences in categorization.*) We define a hypoglycemic event as individual instances where sensor glucose levels drop below or within the specified range. Continuous sequences of hypoglycemic events are termed hypoglycemic episodes.

For each hypoglycemic episode among AID and non-AID users, we compute GV-related metrics before and after the episode, at intervals of ±3, ±6, ±12, ±24, and ±48 hours, to evaluate the impact of varying hypoglycemia levels on GV across different timeframes.

Data are organised into tables (Supplemental Tables A1 to A4) to provide a comprehensive overview of glucose analysis metrics, including TBR, TIR, TAR, SD, HBGI, and LBGI across various time intervals surrounding hypoglycemic episodes. To quantify these metrics, several aforementioned statistical measures are calculated.

## Experimental workflow for individual-level analysis of glycemic variability related to hypoglycemia

A structured, three-staged experimental workflow has been developed (adopted from previous population-level workflow[15]) to examine the effects of hypoglycemia on GV at the individual level [Figure 1].

- **Stage 1 — Data Initialization**: Glucose data is cleaned by removing null values and readings outside 39-400 mg/dL. Hypoglycemia is categorized into three ranges (41-50, 51-60, and 61-70 mg/dL). Time intervals around hypoglycemic events (±3h to ±48h) are established for analysis.
- **Stage 2 — Statistical Analysis**: For each hypoglycemic episode, data is extracted across defined time intervals. Glucose variability metrics are calculated, and episodes are classified by duration (short: <20 min, medium: 20-40 min, long: >40 min).
- **Stage 3 — Visualization**: Glucose variability metrics are plotted over time relative to hypoglycemic episodes, and statistical analyses are performed on pre- and post-hypoglycemia intervals for all individuals.

Due to LBGI's direct correlation with TBR at lower numerical scales[17], we focus on TBR progression plots within this work.

**FIGURE 1:** Three-staged individual-level HypoGV analysis workflow, detailing data

initialisation and preprocessing, statistical analysis across defined time intervals around hypoglycemic episodes, and visualisation of glucose variability progression before and after hypoglycemia.

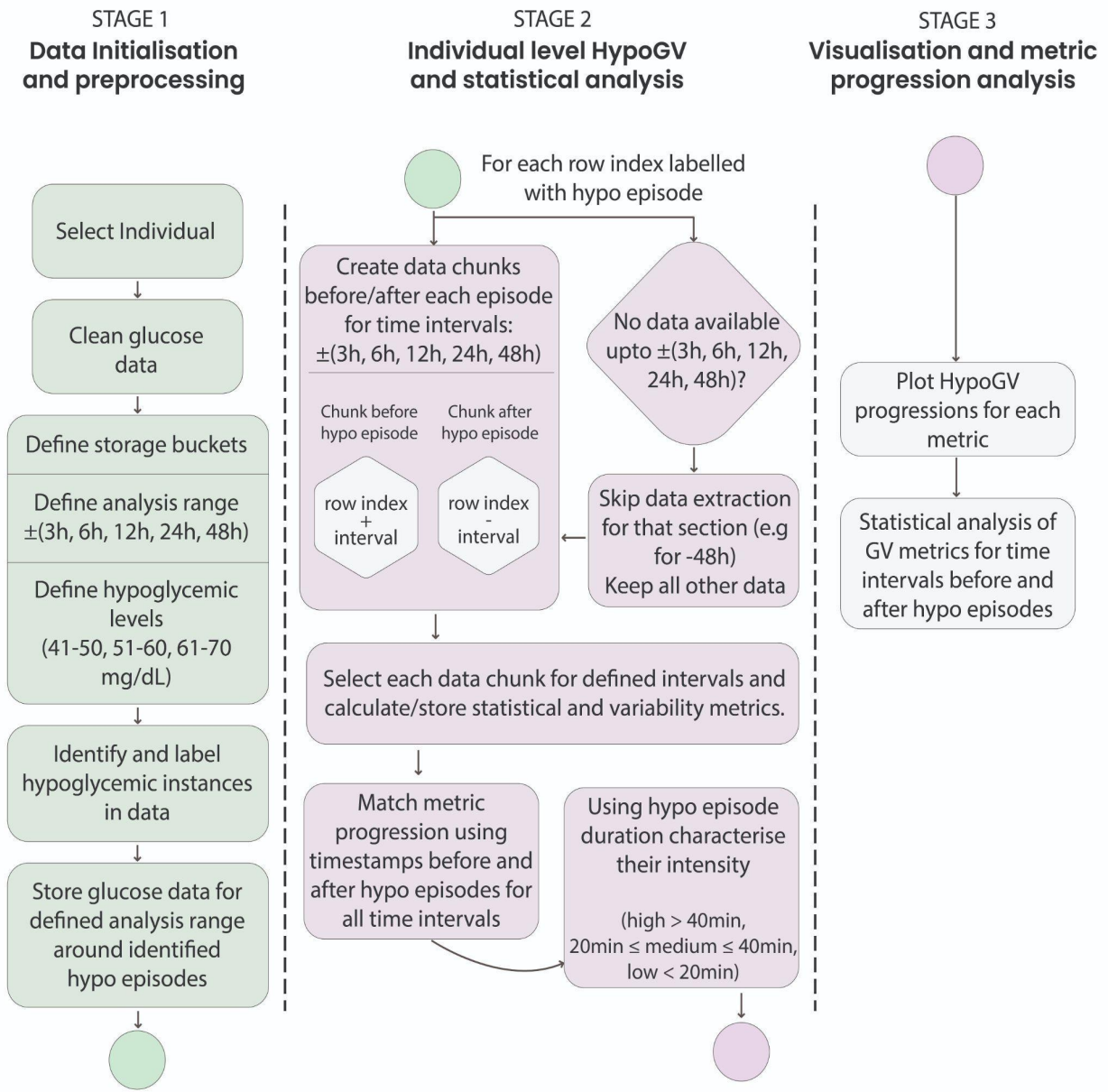

## Results

Our findings highlight significant differences in glycemic outcomes between individuals using AID systems and those relying on non-AID systems, including glycemic variability. AID users consistently demonstrate better glucose outcomes, as evaluated by higher TIR, lower TAR, and reduced TBR. These advantages are reflected across multiple scenarios, including various exercise types. Although mean glucose levels are similar between AID and non-AID users (AID: 143.24mg/dL, non-AID: 144.81mg/dL) ($p > 0.05$), AID users display significantly lower glucose variability, as shown by the reduced standard deviation ($p < 0.05$). Additionally, AID users exhibit less time spent with higher glucose levels, with significant improvements at the 75% quantile ($p < 0.05$). Figure 2 shows the distribution of glucose analysis metrics across AID and non-AID categories, broken out by device/therapy type. Non-AID users show greater variability and suboptimal glycemic outcomes in tighter ranges, despite similar mean glucose outcomes in this dataset.

**FIGURE 2:** Boxplots of Time in Range (TIR), Time Above Range (TAR), and Time Below Range (TBR) for various insulin pump types among AID and non-AID users. The number of users per device type is displayed above each boxplot. AID users generally achieve higher TIR and lower TAR and TBR across most devices.

*Analysis of data shows these users did not appear to use AID during the study period and are thus classified as non-AID

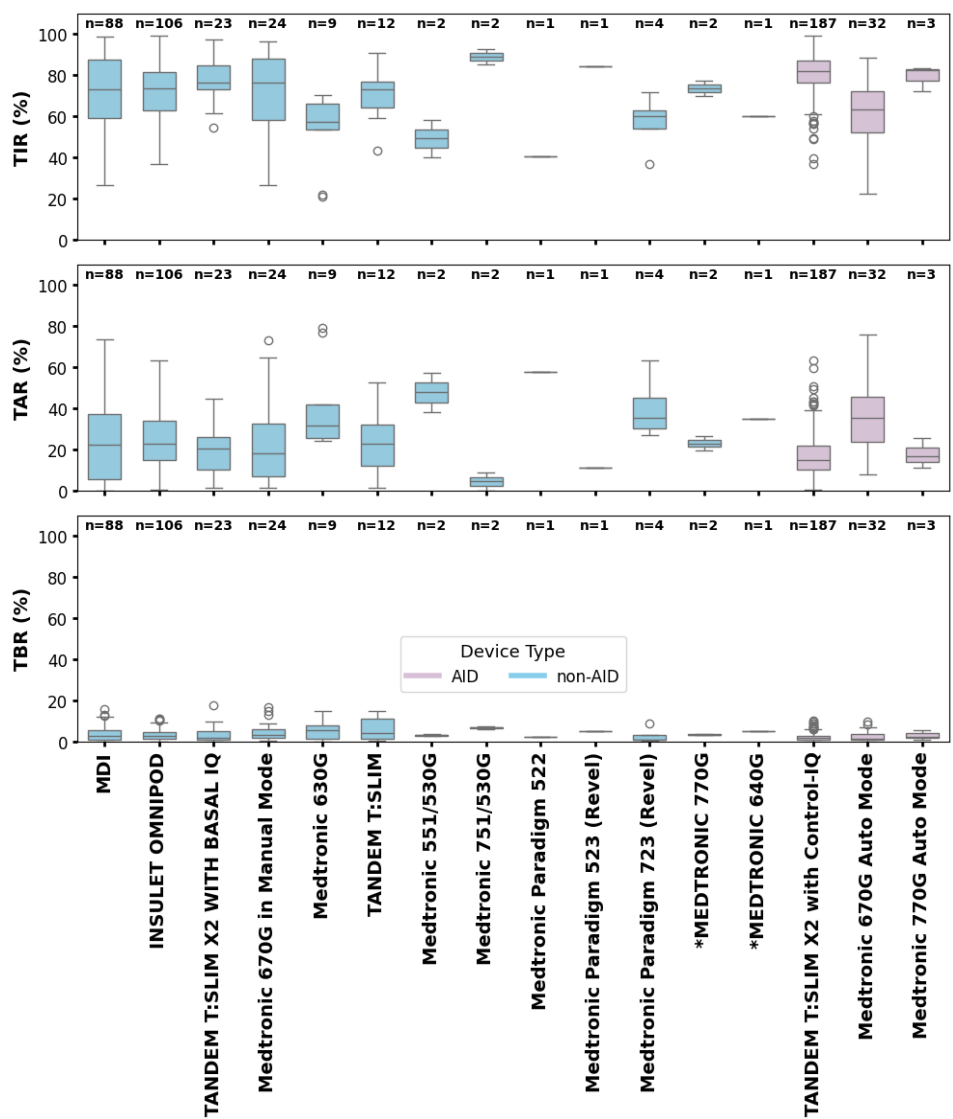

Glucose data for both AID and non-AID users in T1DEXI dataset exhibit non-normal distributions ($p < 0.05$) across all variability metrics (Table A3). In AID data, TIR is negatively skewed, indicating an inclination of the mean towards the left, while other metrics show positive skewness, suggesting values below the mean. Non-AID data follow a similar pattern, indicating comparable distribution characteristics across both groups.

Statistical tests confirm that AID users achieve relatively higher TIR and lower TAR and TBR than non-AID users ($p < 0.05$), although the KS and MWU tests highlight distinct TIR distribution patterns (Table A4).

Figures A2 and A3 show that, while AID outcomes are similar between genders, non-AID males tend to have higher TIR and lower TAR compared to non-AID females, indicating more optimal glycemic outcomes. AID users of both genders overall achieve higher TIR and lower TAR and TBR than non-AID users, with male AID users demonstrating the best outcomes in TIR and TAR and female AID users showing the best results in TBR. Gender-based comparisons (Table 1) indicate that the difference in metrics (TIR, TBR, TAR, SD) between AID and non-AID users was more pronounced among females than males, with females on AID systems demonstrating larger improvements.

**Table 1:** Average values for Mean, SD, TIR (%), TBR (%), and TAR (%) for Males and Females from AID and non-AID groups.

| Sex | Group   | Mean   | SD    | TIR (%) | TAR (%) | TBR (%) |
|-----|---------|--------|-------|---------|---------|---------|
| F   | AID     | 146.03 | 45.93 | 77.25   | 20.68   | 2.07    |
| F   | non-AID | 153.32 | 54.48 | 68.85   | 27.60   | 3.55    |
| M   | AID     | 140.46 | 45.04 | 79.92   | 17.32   | 2.75    |
| M   | non-AID | 136.30 | 46.20 | 77.46   | 17.89   | 4.65    |

Across different exercise types, AID users in T1DEXI achieved more optimal glucose outcomes than non-AID users. In aerobic exercises, AID users achieve a median TIR of 75%, compared to 55% for non-AID users. Among non-AID users, during resistance exercise, TIR varies widely, with an interquartile range (IQR) ranging from 40% to 65%, indicating greater variability in glucose outcomes during these activities. AID users also exhibit lower TAR and TBR, with a median TAR of 12% in interval exercises (compared to 30% for non-AID users) and a median TBR of 3% in aerobic exercises (compared to 10% for non-AID users). Additionally, AID users have lower glucose variability, with an average SD of 43 mg/dL across exercise categories versus 53 mg/dL for non-AID users.

**Population-level analysis of glycemic variability related to hypoglycemia**

Overall, the patterns related to hypoglycemic and glycemic variability observed at the population level in commercial AID systems resemble those in open-source AID systems[15], with GV metrics showing significant fluctuations as hypoglycemia severity increases (from blood glucose level 70 mg/dL to 40 mg/dL). Figure 3 illustrates that Time Below Range (TBR) rises leading up to a hypoglycemic episode, then gradually decreases and stabilises around 48 hours post-episode, and that these progression patterns remain consistent, irrespective of whether the exercise preceded the hypoglycemia or not. Time in Range (TIR) slightly declines before hypoglycemia but returns to stable levels afterwards, while Time Above Range (TAR) shows no specific progression. In Figure A4, Low Blood Glucose Index (LBGI) and High Blood Glucose Index (HBGI) mirror the TBR and TAR patterns, respectively. GV metrics across different hypoglycemia levels do not follow a normal distribution, with statistically significant differences observed across various intervals for both AID and non-AID users (Supplemental Tables A1–A4).

**FIGURE 3:** Population-level comparison of Time in Range (TIR), Time Below Range (TBR), and Time Above Range (TAR) for AID (left column) and non-AID (right column) users, across various time intervals before and after a hypoglycemic episode. Panels (A) and (B) display the distribution of TIR across time, showing a decline before hypoglycemia and recovery afterwards, with AID users maintaining higher TIR percentages overall. Panels (C) and (D) depict the distribution of TBR, which spikes around the hypoglycemic episode, particularly in non-AID users, where TBR remains elevated for longer periods post-episode. Panels (E) and (F) show TAR distributions, with AID users exhibiting generally lower TAR both before and after hypoglycemic episodes, while non-AID users show higher TAR and greater variability across all time intervals.

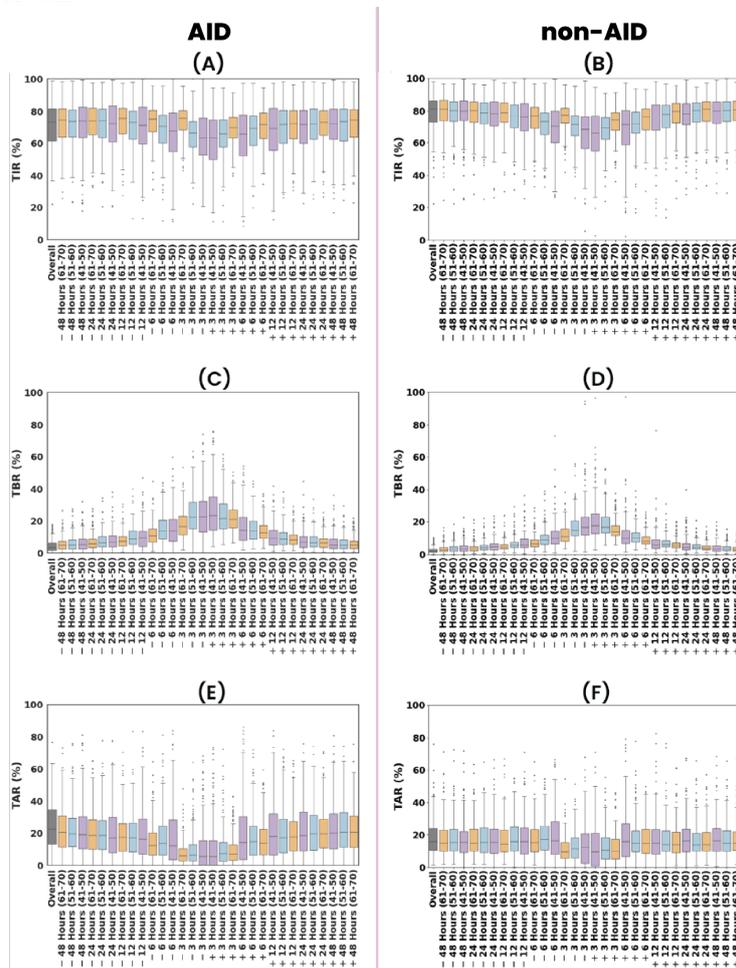

Significant differences (p < 0.05) in TIR, TBR, TAR, HBGI, LBGI, and SD exist between AID and non-AID users for most time intervals (Tables A1 and A2). For TIR, AID users generally maintain higher levels, with similar distributions to non-AID users only at -6h (61–70) (p = 0.227), -6h (41–50) (p = 0.170), and -3h (41-50) (p = 0.171). TBR shows that AID users spend less time below the target range, reducing subsequent hypoglycemia risk. For TAR, several intervals, such as -12h (61–70) (p = 0.288) and -12h (51–60) (p = 0.720), show exceptions in distributions, while most other intervals indicate AID users spend less time above range.

HBGI values indicate mixed results between AID and non-AID users with AID performing similar to non-AID at instances of -24h (61–70) (p = 0.064) and -12h (51–60) (p = 0.613). LBGI shows consistent significant differences across all intervals (p < 0.05), indicating a lower subsequent hypoglycemia risk for AID users around hypoglycemic episodes. For SD, AID compared to non-AID users demonstrate mixed glucose variability across most intervals, with almost half aligning towards AID being better at variability while the other half especially on positive timeframes show the inverse.

**Individual-level analysis of glycemic variability related to hypoglycemia**

The TBR progression for six AID users (Figure 4) and six non-AID users (Figure A5) from -48 to +48 hours around each hypoglycemic episode is visualized. Each line represents progression across one episode, classified by severity based on time duration: short-time duration (< 20 min) is represented with green, medium-time duration (20-40 min) with yellow, and long-time duration (> 40 min) with red. The dominant trends observed in TBR progression among all users show a rising trend as hypoglycemic episodes approach and remain elevated up to 48 hours post-episode before stabilising. Individuals with frequent episodes generally show higher rates of TBR, especially following long-duration episodes (> 40 min), indicating more prolonged recovery.

These figures illustrate that short and medium-duration episodes generally resolve more quickly compared to long-duration events and although AID systems provide some stabilisation, prolonged TBR remains an issue following long-duration episodes of hypoglycemia for all people with diabetes. This highlights the difficulty in achieving glucose stability following long-duration hypoglycemic episodes in both AID and non-AID users.

**FIGURE 4:** TBR progression for six AID users in T1DEXI, segmented by insulin pump type: (A, B) TANDEM T:SLIM X2 WITH CONTROL IQ, (C, D) MEDTRONIC 670G, and (E, F) MEDTRONIC 770G. Long-duration episodes (red) are associated with higher pre- and post-episode TBR values, suggesting prolonged recovery. Medtronic users exhibit greater TBR variability, particularly for long-duration episodes, while Tandem users show more consistent TBR recovery across episode durations.

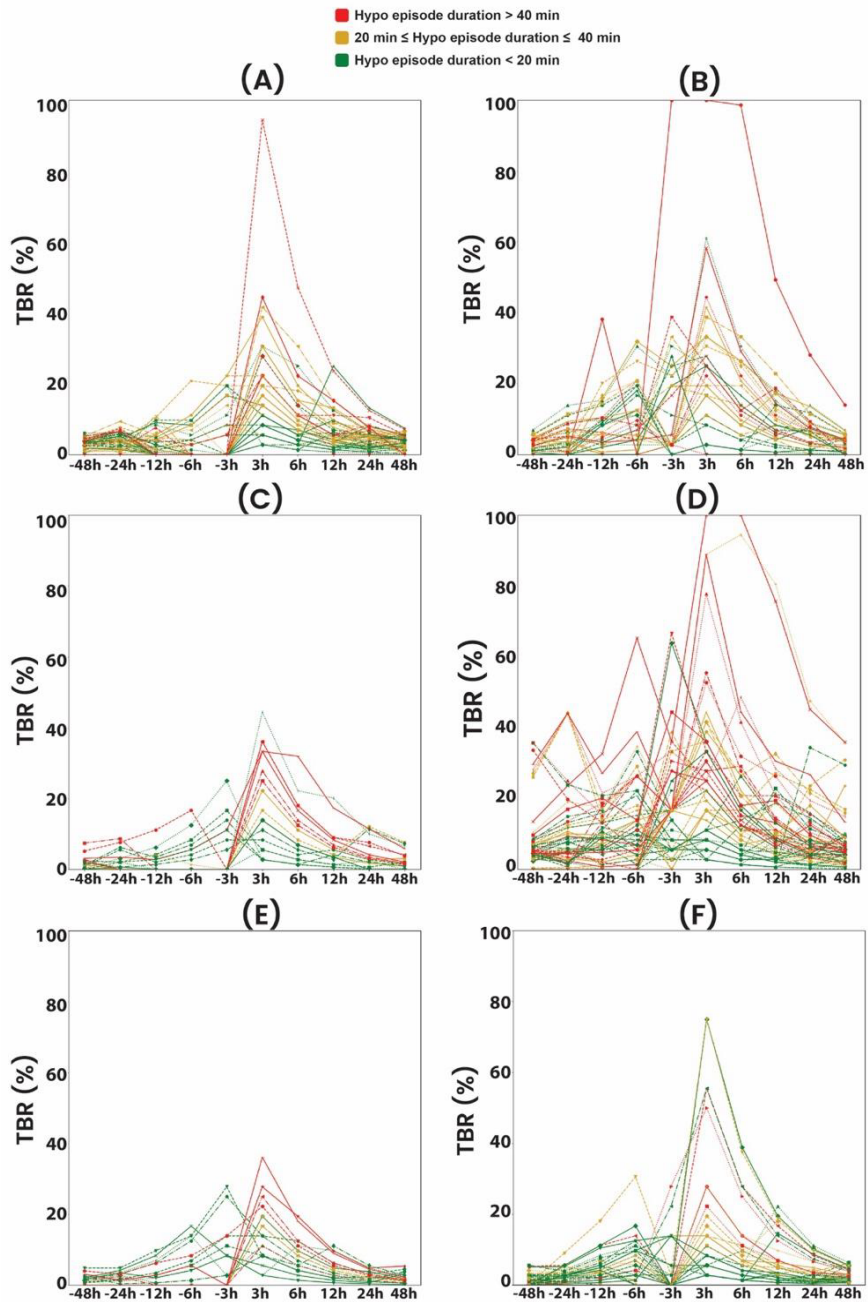

## Discussion

This study aimed to assess pre- and post-exercise glycemic variability among people with type 1 diabetes using a variety of insulin therapy strategies such as AID and non-AID (including MDI and standalone insulin pumps), especially as it relates to instances of hypoglycemia, using the T1DEXI dataset. Analysis of glycemic variability showed that both AID and non-AID users experienced disturbances in GV leading to instances of hypoglycemia, and these patterns persisted regardless of whether the hypoglycemia was related to exercise or not. These disturbances were more pronounced in non-AID users, but AID users still experienced this disruption prior to and following hypoglycemia, with more severe hypoglycemic episodes (41–50 mg/dL) leading to prolonged glucose instability compared to milder episodes (61–70 mg/dL). This trend was observed in both AID and non-AID users, though the magnitude of glucose disturbance was significantly greater in non-AID users, suggesting that while AID systems mitigate some of the glycemic burdens, they do not entirely prevent prolonged glucose instability following severe hypoglycemia. These findings confirm that glycemic variability evaluation of timeseries data of CGM may be beneficial in both developing additional strategies to reduce subsequent hypoglycemia as well as iterating on diabetes technology to further dampen patterns of hypoglycemia for people with type 1 diabetes.

While glycemic variability is often studied, it is only recently being increasingly used with timeseries data for CGM and often looked at a broad time period such as 14 days or 30 days[18]. Previous analyses with T1DEXI have focused on time frames up to 24 hours but in relationship with exercise sessions[19] rather than that of hypoglycemia. In this work, we observed utility in looking at specific time windows before and after an instance of hypoglycemia, as we observed that disturbances in GV usually normalize within 48 hours following hypoglycemia, though this stabilization period varied depending on the severity and duration of the hypoglycemic episode. Long-duration hypoglycemia events (>40 min) resulted in significantly prolonged TBR elevation, suggesting that sustained low glucose exposure may contribute to slower metabolic recovery, possibly due to increased counter-regulatory hormonal responses or delayed glycogen replenishment. This was true for both AID and non-AID users who exhibited similar patterns with different amplitudes and did not vary significantly by exercise type, nor whether the hypoglycemia was related to exercise or not.

Interestingly, one sub-group difference that we did observe was that males overall tended to have higher TIR, and lower TAR compared to females, illustrating slightly improved glycemic outcomes comparatively. This of course could be influenced by bias within the study population but is interesting in comparison to other GV analyses in other populations such as the one in open-source AID users that found a similar result[20]. We also observed a larger difference between females using AID and non-AID as compared to the difference in males using AID and non-AID as shown in Table 1. Females using AID demonstrated more TIR (12.2% difference), less TBR (41.69% difference), less TAR (25.07% difference) and SD (15.69% difference) compared to females using non-AID. This difference is less in males using AID compared to males using non-AID (ΔTIR:3.18%, ΔTAR:3.19%, ΔTBR:40.86%, ΔSD:2.51%)), as it is between the two female groups, suggesting that AID may offer greater relative benefits for female users in glucose stability compared to males. This difference could be partially influenced by hormonal fluctuations related to the menstrual cycle, as previous research suggests that estrogenic and progesterone variations impact insulin sensitivity and glycemic variability[21]. The lack of menstrual cycle tracking in this dataset prevents further confirmation, but future studies should investigate whether cycle-aware insulin adjustments could further optimise AID outcomes in some female users.

Although the users studied here in AID and non-AID groups are independent cohorts, this suggests that AID is likely to help female users in T1D management more, compared to males using the same, and further studies of gender aspects of the T1DEXI and other AID-related datasets are warranted. Additionally, gender-based differences in hypoglycemia recovery post-exercise should be explored further, as previous work suggests that females may experience prolonged post-exercise insulin sensitivity compared to males, potentially contributing to greater post-hypoglycemia glucose instability in non-AID females. This analysis is currently without regard to the menstrual cycle, which is known to affect some individuals more than others, so there are even possibly larger gains for sub-groups of females with T1D to benefit from AID use compared to other therapy modalities. This suggests a need for a future meta-analysis across studies and datasets to better understand gender dynamics in glycemic outcomes and perhaps evaluate the menstrual cycle as it relates to glycemic outcomes. It may also be determined in the future that it is inter-individual variability, rather than gender differences, that make the largest difference in outcomes, but this will only be able to be evaluated in the future as we have more detailed datasets that include user interaction data such as adjusting targets and profile changes as well as full insight into insulin delivery decisions.

These findings further build on previous research, including a previous study with similar methods evaluating patterns of GV related to hypoglycemia in n=122 open-source AID users[20]. The patterns observed in population-based data for glycemic variability disturbances before and after hypoglycemia persist in both datasets, which shows that this finding is not limited to certain brands or types of AID systems, whether open source or commercial, and points to an opportunity to fold this into AID systems or other diabetes technologies as an additional factor for reducing recurring hypoglycemic, particularly by incorporating GV trend monitoring as a predictive variable within automated insulin delivery algorithms. Future iterations of AID systems could integrate rolling 48-hour GV trend analysis to preemptively adjust insulin dosing in users with persistent GV fluctuations preceding hypoglycemia, potentially reducing the frequency of recurring low glucose events.

While the previous study evaluated population-level outcomes in OS-AID users, exercise data was not available as it is in the T1DEXI dataset, so we encourage replication in other datasets in the future that also have the exercise data available, to assess whether the finding that exercise type did not change the glucose variability patterns persists and is a scalable observation to the entire population of people with T1D.

**Limitations**

As we have noted, there are limitations to the study, such as the lack of consistent carbohydrate or meal data throughout the entire T1DEXI dataset, although some limited data on correction carbohydrates is available. As such, we are unable to analyse user actions such as strategies for eating carbs prior to exercise, or the timing of such, as well as the influence of correction or rescue carbohydrate intake relative to all instances of hypoglycemia present in the dataset. We encourage future dataset generation from studies to include this information for the entire period of the CGM data collected, as it would be beneficial to evaluate user action along with insulin delivery data and CGM data to understand the interaction between humans and diabetes devices to determine what interventions may be most beneficial for reducing hypoglycemia. The gender-related analyses should also be replicated with future datasets that include menstrual cycle-related data to help identify whether intra-individual or menstrual cycle timing and gender play larger roles in the variable outcomes.

**Conclusion**

Mirroring previous research broadly as well as specific to the T1DEXI dataset, we find that AID users experience improved glycemic variability outcomes compared to non-AID users, but more specifically that both AID and non-AID users experience similar patterns of glycemic variability disturbances preceding and following instances of hypoglycemic up to 48 hours later. This suggests additional studies with glycemic variability for specific periods such as before and after exercise or hypoglycemia may provide pointers to future improvements for AID systems or other diabetes technology to further reduce the occurrences of hypoglycemia, especially as it relates to exercise. This is a key area of opportunity for further advancements in AID systems and other diabetes technologies, such as using glycemic variability as a factor in algorithmic or human decision-making, to optimize outcomes and minimize the risk of hypoglycemia during or after exercise.

**Abbreviations:** AID (Automated Insulin Delivery); CGM (Continuous Glucose Monitoring); GV (glycemic or glucose variability); high blood glucose index (HBGI); Kolmogorov-Smirnov (KS); Shapiro-Wilk (SW); Mann-Whiney U (MWU); low blood glucose index (LBGI); glucose management index (GMI); coefficient of variation (CV); Mann-Whitney (MS); Standard Deviation (SD); T1D (Type 1 Diabetes); TIR (Time In Range); TBR (time before range); TAR (time after range)

**Figures and Tables:** 4 figures and 1 table within the manuscript (an additional 5 figures and 4 tables are in the supplemental appendix for a total of 9 figures and 5 tables overall)

**Author Contributions:** Conceptualization, A.Z., A.So., D.M.L., A.Sh.; methodology, A.Z., A.So., D.M.L., A.Sh.; software, A.Z., A.So., D.M.L., A.Sh.; validation, A.Z., A.So., D.M.L., A.Sh.; formal analysis, A.Z., A.So., D.M.L., A.Sh.; investigation, D.M.L. and A.Sh.; resources, D.M.L. and A.S.; data curation, A.Z., A.So., D.M.L., A.Sh.; writing—original draft preparation, A.Z., D.M.L., A.Sh.; writing—review and editing, A.Z., A.So., D.M.L., A.Sh.; visualization, A.Z., A.So., D.M.L., A.Sh.; supervision, D.M.L. and A.Sh.; project administration, A.Sh.; funding acquisition, D.M.L. and A.Sh. All authors have read and agreed to the published version of the manuscript.

**Conflicts of Interest:** All authors declare no financial conflict of interest. The funders had no role in the design of the study; in the analyses, or interpretation of data; in the writing of the manuscript; or in the decision to publish the results.

**Institutional Review Board Statement:** Ethics approval is granted by UCD's Human Research Ethics Committee – Sciences (HREC-LS) with reference number **LS-LRSD-23-264-Shahid** which meets the criteria for a low-risk study involving secondary data.

**Data Availability Statement:** All programming scripts and tools developed for the analysis in this paper are made public and online. All data comes from the T1DEXI dataset.

**Funding:** This work was performed under a grant from the Leona M. and Harry B. Helmsley Charitable Trust (Grant #: 2407-07175).

# SUPPLEMENTARY APPENDIX

**FIGURE A1:** Density plot of glucose data collection days for AID and non-AID users in T1DEXI. Both groups exhibit similar data distributions, with median values around 27 days. AID users' data has a slightly broader range (4.49 to 54.74 days) compared to non-AID users (1.94 to 47.11 days), though mean values remain close (26.64 days for AID, 26.29 days for non-AID).

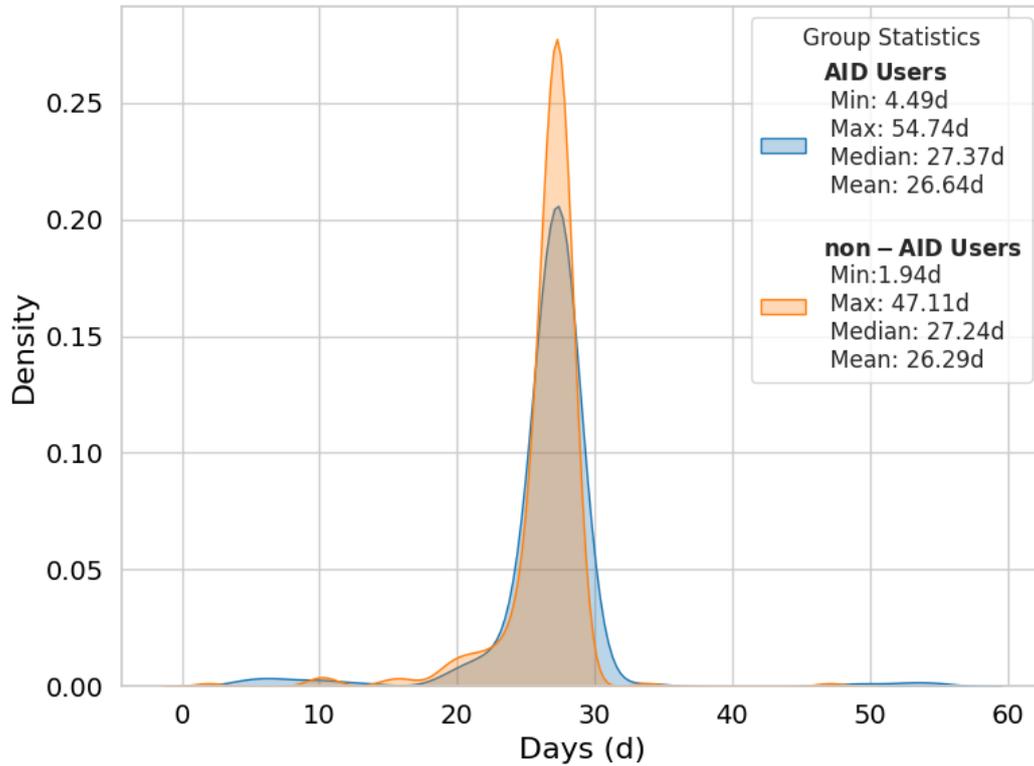

**FIGURE A2:** Gender-wise comparison of TIR, TAR, TBR, SD, and Mean Glucose for AID and non-AID users. Across both AID and non-AID groups, females exhibit slightly higher TIR, and lower TAR compared to males, suggesting better glycemic outcomes. AID users, irrespective of gender, achieve higher TIR and lower TAR and TBR than non-AID users. Additionally, females tend to have lower TBR and SD than males in both AID and non-AID groups, with female AID users achieving the best overall outcomes.

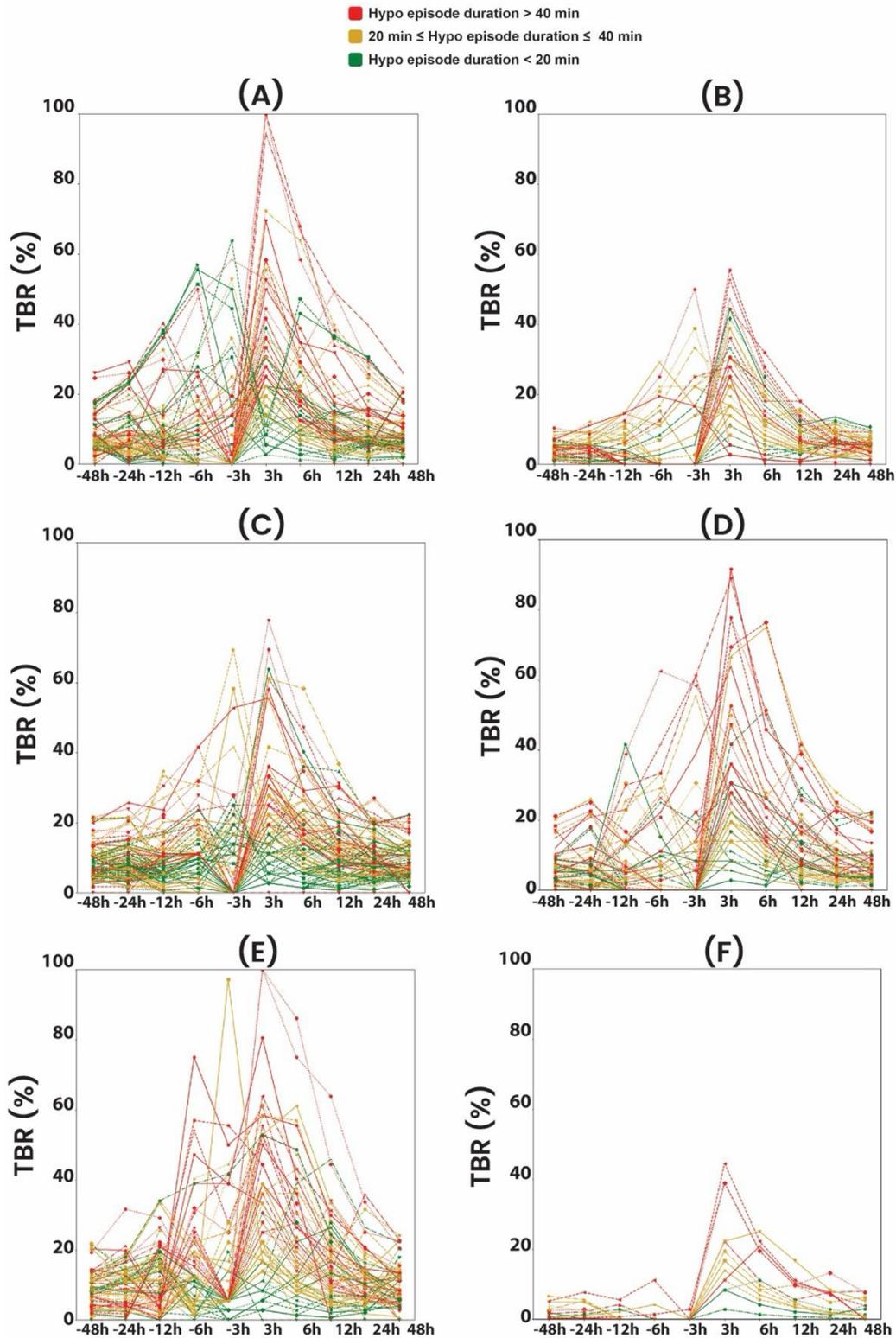

**FIGURE A3:** Activity-wise comparison of TIR, TAR, TBR, SD, and Mean Glucose for AID and non-AID users across different exercise types: Resistance, Interval, and Aerobic. AID users generally achieve higher TIR and lower TAR and TBR across all exercise types compared to non-AID users. Panel (A) shows that AID users attain higher TIR, particularly during interval and aerobic exercises. Panels (B) and (C) illustrate that TAR and TBR are lower for AID users across all exercise types, with non-AID users displaying greater variability. Panels (D) and (E) demonstrate that AID users sustain lower SD and mean glucose levels, especially during aerobic exercises. Overall, aerobic exercises yield the highest TIR and lowest TAR for both AID and non-AID users.

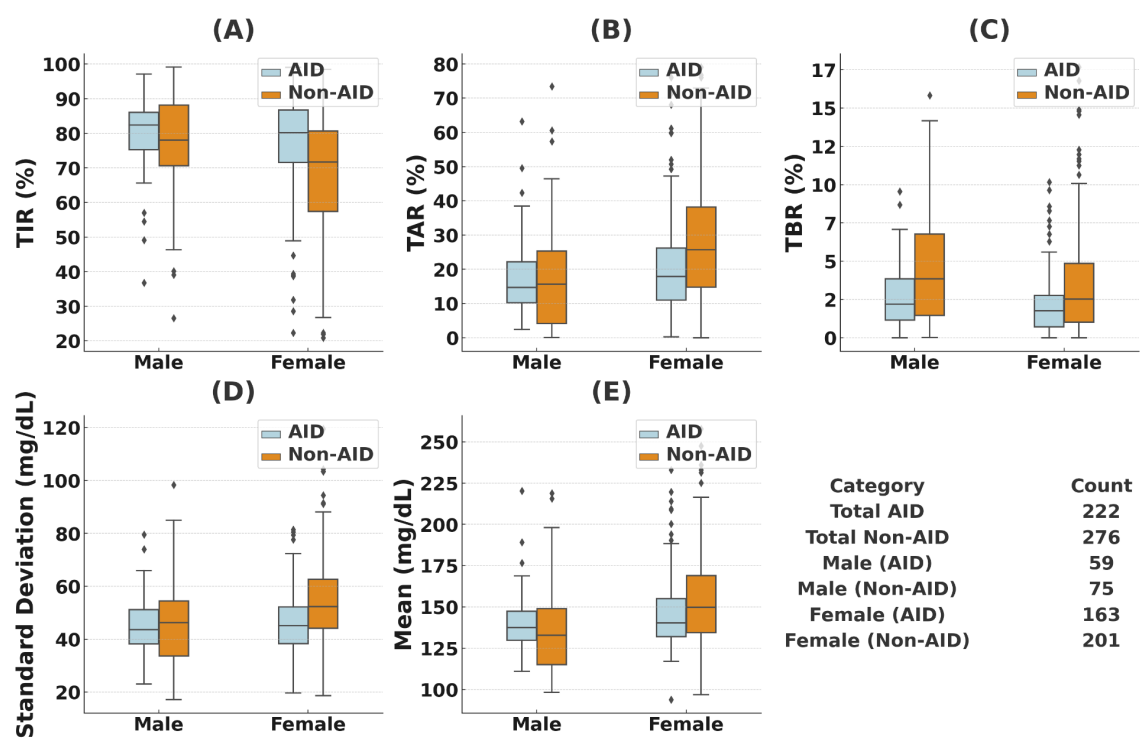

**FIGURE A4:** Population-level comparison of High Blood Glucose Index (HBGI), Standard Deviation (SD), and Low Blood Glucose Index (LBGI) for AID (Panels A, C, E) and non-AID (Panels B, D, F) users across various time intervals before and after a hypoglycemic episode. Panels (A) and (B) display the distribution of HBGI, showing generally lower values for AID users across time intervals, with non-AID users having higher and more variable HBGI. Panels (C) and (D) compare SD, indicating more stable glucose outcomes in AID users, with lower overall variability compared to non-AID users, particularly around hypoglycemic episodes. Panels (E) and (F) show LBGI distribution, where AID users exhibit lower LBGI, especially during and after hypoglycemic episodes, whereas non-AID users have higher LBGI and greater variability in the same periods.

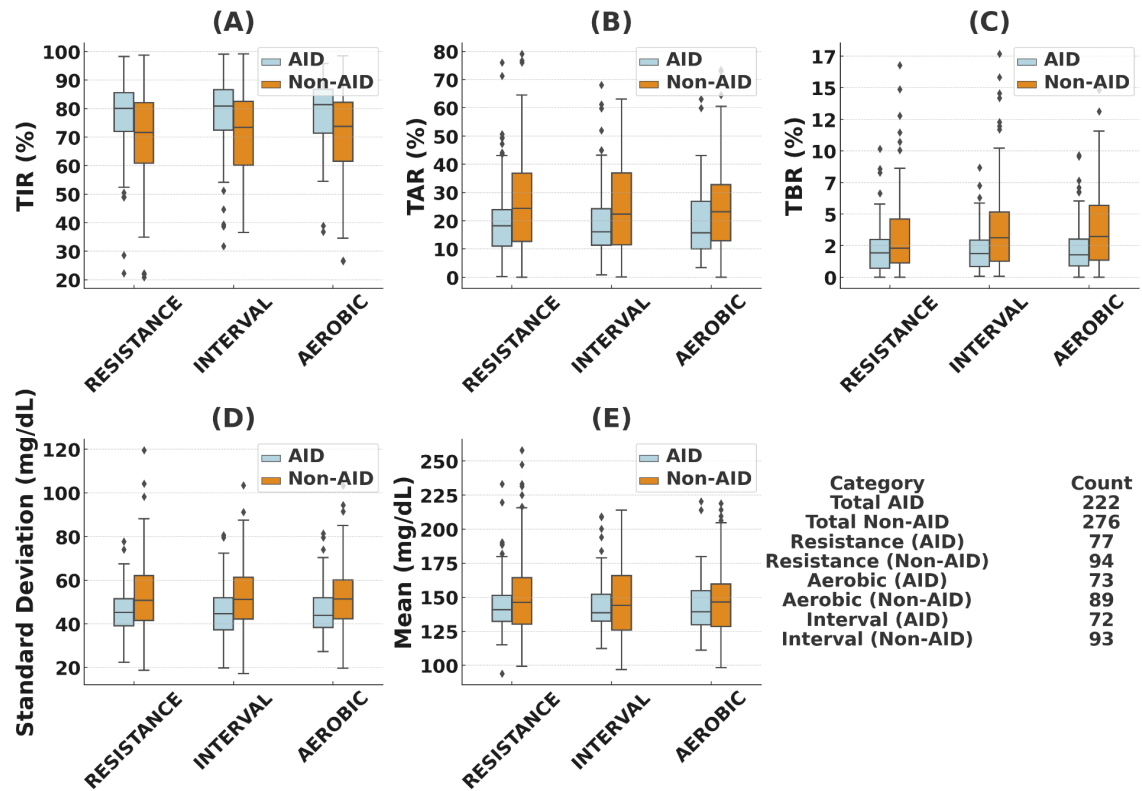

**FIGURE A5:** TBR progression for six non-AID users in T1DEXI, segmented by insulin delivery method: (A, B) INSULET OMNIPOD, (C, D) TANDEM T:SLIM X2 BASAL IQ, and (E, F) MDI (Multiple Daily Injections). Long duration episodes (red) result in higher post-episode TBR, with MDI users showing the highest variability. OMNIPOD and Tandem BASAL IQ users exhibit more consistent but elevated TBR for longer durations.

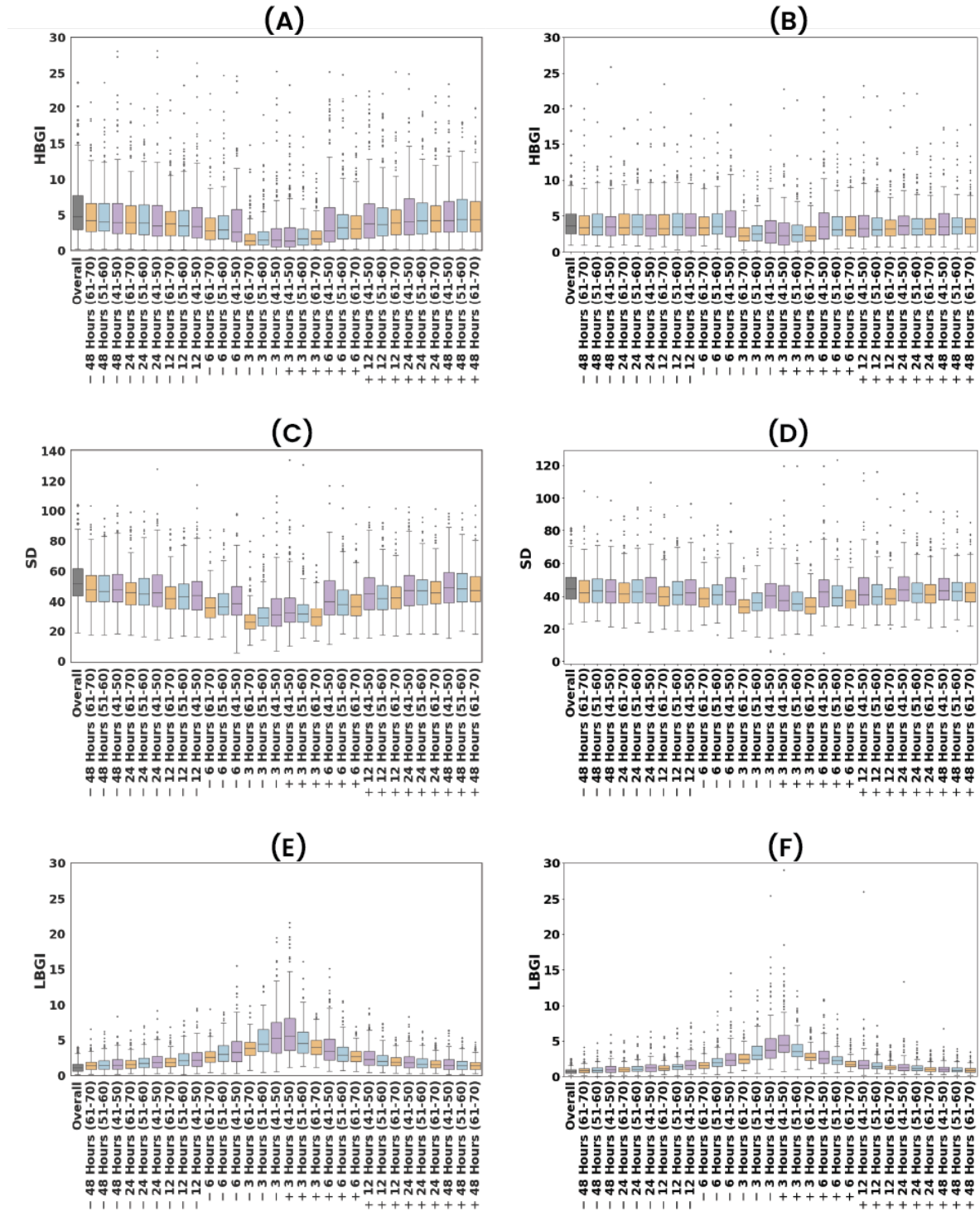

**Table A1:** Population level comparison of glucose analysis metrics distributions between AID and non-AID users across Time in Range (TIR), Time Below Range (TBR), and Time Above Range (TAR) using Z-test and Kolmogorov–Smirnov (KS) test for various time intervals surrounding hypoglycemic episodes. The KS-test generally aligns with the Z-test findings, confirming significant distribution differences in most intervals across the three metrics.

| Time Interval | Z-test (TIR) | KS-test (TIR) | Z-test (TBR) | KS-test (TBR) | Z-test (TAR) | KS-test (TAR) |
|---|---|---|---|---|---|---|
| Overall | <0.05 | <0.05 | <0.05 | <0.05 | <0.05 | <0.05 |
| -48h (61-70) | <0.05 | <0.05 | <0.05 | <0.05 | <0.05 | <0.05 |
| -48h (51-60) | <0.05 | <0.05 | <0.05 | <0.05 | <0.05 | <0.05 |
| -48h (41-50) | <0.05 | <0.05 | <0.05 | <0.05 | <0.05 | <0.05 |
| -24h (61-70) | <0.05 | <0.05 | <0.05 | <0.05 | 0.081 | 0.165 |
| -24h (51-60) | <0.05 | <0.05 | <0.05 | <0.05 | 0.082 | 0.090 |
| -24h (41-50) | <0.05 | <0.05 | <0.05 | <0.05 | 0.248 | 0.139 |
| -12h (61-70) | <0.05 | <0.05 | <0.05 | <0.05 | 0.288 | 0.090 |
| -12h (51-60) | <0.05 | <0.05 | <0.05 | <0.05 | 0.720 | 0.139 |
| -12h (41-50) | <0.05 | <0.05 | <0.05 | <0.05 | 0.248 | 0.139 |
| -6h (61-70) | 0.227 | 0.233 | <0.05 | <0.05 | <0.05 | <0.05 |
| -6h (51-60) | <0.05 | <0.05 | <0.05 | <0.05 | 0.119 | <0.05 |
| -6h (41-50) | 0.170 | 0.271 | <0.05 | <0.05 | 0.366 | <0.05 |
| -3h (61-70) | <0.05 | 0.100 | <0.05 | <0.05 | <0.05 | <0.05 |
| -3h (51-60) | <0.05 | <0.05 | <0.05 | <0.05 | <0.05 | <0.05 |
| -3h (41-50) | 0.171 | <0.05 | <0.05 | <0.05 | <0.05 | <0.05 |
| +3h (61-70) | <0.05 | <0.05 | <0.05 | <0.05 | 0.314 | <0.05 |
| +3h (51-60) | <0.05 | 0.072 | <0.05 | <0.05 | <0.05 | <0.05 |
| +3h (41-50) | <0.05 | <0.05 | <0.05 | <0.05 | <0.05 | <0.05 |
| +6h (61-70) | <0.05 | <0.05 | <0.05 | <0.05 | 0.685 | 0.217 |
| +6h (51-60) | <0.05 | <0.05 | <0.05 | <0.05 | 0.893 | <0.05 |
| +6h (41-50) | <0.05 | <0.05 | <0.05 | <0.05 | 0.793 | 0.102 |
| +12h (61-70) | <0.05 | <0.05 | <0.05 | <0.05 | <0.05 | <0.05 |
| +12h (51-60) | <0.05 | <0.05 | <0.05 | <0.05 | <0.05 | <0.05 |
| +12h (41-50) | <0.05 | <0.05 | <0.05 | <0.05 | <0.05 | <0.05 |
| +24h (61-70) | <0.05 | <0.05 | <0.05 | <0.05 | <0.05 | <0.05 |
| +24h (51-60) | <0.05 | <0.05 | <0.05 | <0.05 | <0.05 | <0.05 |
| +24h (41-50) | <0.05 | <0.05 | <0.05 | <0.05 | <0.05 | <0.05 |

**Table A2:** Population level comparison of glucose variability metrics distributions between AID and non-AID users across High Blood Glucose Index (HBGI), Low Blood Glucose Index (LBGI), and Standard Deviation (SD) using Z-test and Kolmogorov–Smirnov (KS) test for various time intervals surrounding hypoglycemic episodes. The KS-test results align closely with the Z-test findings, confirming significant distribution differences in most intervals for HBGI, LBGI, and SD.

| Time Interval | Z-test (HBGI) | KS-test (HBGI) | Z-test (LBGI) | KS-test (LBGI) | Z-test (SD) | KS-test (SD) |
|---|---|---|---|---|---|---|
| Overall | <0.05 | <0.05 | <0.05 | <0.05 | <0.05 | <0.05 |
| -48h (61-70) | <0.05 | <0.05 | <0.05 | <0.05 | <0.05 | <0.05 |
| -48h (51-60) | <0.05 | <0.05 | <0.05 | <0.05 | <0.05 | <0.05 |
| -48h (41-50) | <0.05 | <0.05 | <0.05 | <0.05 | <0.05 | <0.05 |
| -24h (61-70) | 0.064 | 0.061 | <0.05 | <0.05 | <0.05 | <0.05 |
| -24h (51-60) | 0.127 | 0.429 | <0.05 | <0.05 | <0.05 | 0.098 |
| -24h (41-50) | <0.05 | 0.096 | <0.05 | <0.05 | <0.05 | <0.05 |
| -12h (61-70) | 0.470 | 0.204 | <0.05 | <0.05 | 0.187 | 0.096 |
| -12h (51-60) | 0.613 | 0.153 | <0.05 | <0.05 | 0.451 | 0.336 |
| -12h (41-50) | 0.206 | 0.074 | <0.05 | <0.05 | 0.278 | 0.324 |
| -6 h (61-70) | <0.05 | <0.05 | <0.05 | <0.05 | <0.05 | <0.05 |
| -6 h (51-60) | 0.320 | <0.05 | <0.05 | <0.05 | 0.051 | <0.05 |
| -6 h (41-50) | 0.557 | <0.05 | <0.05 | <0.05 | 0.051 | <0.05 |
| -3 h (61-70) | <0.05 | <0.05 | <0.05 | <0.05 | <0.05 | <0.05 |
| -3 h (51-60) | <0.05 | <0.05 | <0.05 | <0.05 | <0.05 | <0.05 |
| -3 h (41-50) | 0.171 | <0.05 | <0.05 | <0.05 | <0.05 | <0.05 |
| +3 h (61-70) | 0.444 | <0.05 | <0.05 | <0.05 | <0.05 | <0.05 |
| +3 h (51-60) | 0.083 | <0.05 | <0.05 | <0.05 | <0.05 | <0.05 |
| +3 h (41-50) | 0.092 | <0.05 | <0.05 | <0.05 | <0.05 | <0.05 |
| +6 h (61-70) | 0.994 | 0.092 | <0.05 | <0.05 | 0.538 | 0.327 |
| +6 h (51-60) | 0.907 | 0.058 | <0.05 | <0.05 | 0.264 | <0.05 |
| +6 h (41-50) | 0.994 | 0.059 | <0.05 | <0.05 | 0.421 | <0.05 |
| +12h (61-70) | 0.075 | <0.05 | <0.05 | <0.05 | 0.111 | <0.05 |
| +12h (51-60) | <0.05 | <0.05 | <0.05 | <0.05 | 0.187 | <0.05 |
| +12h (41-50) | <0.05 | <0.05 | <0.05 | <0.05 | 0.278 | <0.05 |
| +24h (61-70) | <0.05 | <0.05 | <0.05 | <0.05 | 0.084 | <0.05 |
| +24h (51-60) | <0.05 | <0.05 | <0.05 | <0.05 | 0.069 | 0.096 |
| +24h (41-50) | <0.05 | <0.05 | <0.05 | <0.05 | 0.397 | 0.121 |
| +48h (61-70) | <0.05 | <0.05 | <0.05 | <0.05 | <0.05 | <0.05 |
| +48h (51-60) | <0.05 | <0.05 | <0.05 | <0.05 | 0.263 | 0.277 |
| +48h (41-50) | <0.05 | <0.05 | <0.05 | <0.05 | 0.175 | 0.333 |

**Table A3:** Shapiro-Wilk (SW) test results and skewness scores across various metrics for AID and non-AID data. The table presents the Shapiro-Wilk test results and skewness scores for various metrics related to glucose variability, comparing AID and non-AID users. Both datasets show non-normal distributions (p < 0.05) across all metrics. For Time in Range (TIR), the skewness score is more negative in AID data (-1.32) compared to non-AID (-0.64), suggesting a heavier tail on the lower end in AID users. The skewness score for Time Above Range (TAR) is higher for AID users (1.39) than for non-AID users (0.72), indicating a stronger skew towards higher values in AID users. For Time Below Range (TBR), both groups exhibit positive skewness, but AID users show a higher skewness score (1.67) compared to non-AID users (1.48), implying a stronger skew towards low blood glucose levels in AID users. Across the mean, standard deviation, and percentiles (25\%, 50\%, and 75\%), both groups exhibit similar trends of positive skewness, with AID data tending to have slightly higher skewness for certain metrics.

| Metric | AID Data (SW W) | AID Data (SW p-value) | AID Data (Skewness) | Non-AID Data (SW W) | Non-AID Data (SW p-value) | Non-AID Data (Skewness) |
|---|---|---|---|---|---|---|
| TIR | 0.90 | <0.05 | -1.32 | 0.97 | <0.05 | -0.64 |
| TAR | 0.89 | <0.05 | 1.39 | 0.95 | <0.05 | 0.72 |
| TBR | 0.85 | <0.05 | 1.67 | 0.86 | <0.05 | 1.48 |
| Mean | 0.90 | <0.05 | 1.30 | 0.96 | <0.05 | 0.85 |
| SD | 0.97 | <0.05 | 0.66 | 0.97 | <0.05 | 0.70 |
| 25% | 0.90 | <0.05 | 1.46 | 0.93 | <0.05 | 1.22 |
| 50% | 0.88 | <0.05 | 1.45 | 0.94 | <0.05 | 1.05 |
| 75% | 0.90 | <0.05 | 1.30 | 0.94 | <0.05 | 1.07 |

**Table A4:** Z-test, Kolmogorov-Smirnov (KS) test, and Mann-Whitney U (MWU) test results for TIR, TAR, TBR, Mean, SD, and percentiles (25%, 50%, 75%) comparing AID and non-AID users. For TIR, TAR, TBR, SD, and the 75th percentile, all three tests show statistical differences ($p < 0.05$), indicating significant variability between the AID and non-AID groups in these metrics. The mean and 25th percentile show mixed results: while the Z-test and KS-test reveal no significant difference ($p > 0.05$) for the mean, the MWU-test for the 25th percentile indicates significance ($p < 0.05$). The 50th percentile shows significant differences in the Z-test and KS-test, but the MWU-test does not.

|        | Test     | Statistic | p-value  |
|--------|----------|-----------|----------|
| **TIR**    | Z-test   | 5.05      | < 0.05   |
|        | KS-test  | 0.27      | < 0.05   |
|        | MWU-test | 38792.50  | < 0.05   |
| **TAR**    | Z-test   | -3.74     | < 0.05   |
|        | KS-test  | 0.23      | < 0.05   |
|        | MWU-test | 24971.50  | < 0.05   |
| **TBR**    | Z-test   | -5.90     | < 0.05   |
|        | KS-test  | 0.26      | < 0.05   |
|        | MWU-test | 22866.50  | < 0.05   |
| **Mean**   | Z-test   | -1.83     | > 0.05   |
|        | KS-test  | 0.16      | > 0.05   |
|        | MWU-test | 27931.50  | > 0.05   |
| **SD**     | Z-test   | -4.85     | < 0.05   |
|        | KS-test  | 0.23      | < 0.05   |
|        | MWU-test | 22946.50  | < 0.05   |
| **25%**    | Z-test   | 1.74      | > 0.05   |
|        | KS-test  | 0.24      | < 0.05   |
|        | MWU-test | 35156.50  | < 0.05   |
| **50%**    | Z-test   | -2.20     | < 0.05   |
|        | KS-test  | 0.17      | < 0.05   |
|        | MWU-test | 27671.50  | > 0.05   |
| **75%**    | Z-test   | -3.36     | < 0.05   |
|        | KS-test  | 0.21      | < 0.05   |
|        | MWU-test | 25420.50  | < 0.05   |